\documentclass[aps,twocolumn]{revtex4}%
\usepackage{amssymb}
\usepackage{amsmath}
\usepackage{graphicx}%
\usepackage{amsfonts}

\begin{document}
\preprint{Draft C 9-4-2003}
\title{Direct Observation of Dynamical Switching between Two Driven Oscillation States of a
Josephson Junction.}
\author{I. Siddiqi, R. Vijay, F. Pierre, C.M. Wilson, L. Frunzio, M. Metcalfe, C.
Rigetti, R.J. Schoelkopf, and M.H. Devoret}
\affiliation{Departments of Applied Physics and Physics, Yale University, New Haven,
Connecticut 06520-8284}
\author{D. Vion and D. Esteve}
\affiliation{Service de Physique de l'Etat Condens\'{e}, CEA-Saclay, F-91191 Gif-sur-Yvette (France)}

\begin{abstract}
We performed a novel phase sensitive microwave reflection
experiment which directly probes the dynamics of the Josephson
plasma resonance in both the linear and non-linear regime. When
the junction was driven below the plasma frequency into the
non-linear regime, we observed for the first time the transition
between two different dynamical states predicted for non-linear
systems. In our experiment, this transition appears as an abrupt
change in the reflected signal phase at a critical excitation
power.

\end{abstract}
\volumeyear{year}
\volumenumber{number}
\issuenumber{number}
\eid{identifier}
\received[Received text]{date}

\revised[Revised text]{date}

\accepted[Accepted text]{date}

\published[Published text]{date}

\startpage{1}
\endpage{2}
\maketitle

As first understood by Josephson, a superconducting tunnel
junction can be viewed as a non-linear electrodynamic oscillator
\cite{Josephson}. The tunneling of Cooper pairs manifests itself
as a non-linear inductance that shunts the linear junction
self-capacitance $C_{J}$, formed by the junction electrodes and
the tunnel oxide layer. The constitutive relation of the
non-linear inductor can be written as $I(t)=I_{0}\sin\delta\left(
t\right) $, where $I(t)$, $\delta\left( t\right)
=\int_{-\infty}^{t}dt' V(t')/\varphi_{0}$ and $V(t)$ are the
current, gauge-invariant phase-difference and voltage
corresponding to the inductor, respectively, while the parameter
$I_{0}$ is the junction critical current. Here
$\varphi_{0}=\hbar/2e$ is the reduced flux quantum. For small
oscillation amplitude, the frequency of oscillation is given for
zero bias current by the so-called plasma frequency $\omega
_{P}=1/\sqrt{L_{J}C_{J}}$ where $L_{J}=\varphi_{0}/I_{0}$ is the
effective junction inductance. As the oscillation amplitude
increases, the oscillation frequency decreases, an effect which
has been measured in both the classical and quantum regime
 \cite{Dahm,Pedersen,Devoret,Martinis,Yurke1,Holst}. However, a more
dramatic non-linear effect should manifest itself if the junction
is driven with an AC current $i\sin\omega t$ at a frequency
$\omega$ slightly below $\omega_{P}$. If the quality factor
$Q=C_{J}\omega _{P}/Re[Z^{-1}(\omega _{P})]$ is greater than
$\sqrt{3}/2\alpha$, where $Z(\omega _{P})$ is impedance of the
junction electrodynamic environment and
$\alpha=1-\omega/\omega_{P}$ the detuning parameter, then the
junction should switch from one dynamical oscillation state to
another when $i$ is ramped above a critical value $i_{c1}$
\cite{Dykman}. For $i<i_{c1}$, the oscillation state would be
low-amplitude and phase-lagging while for $i>i_{c1}$, the
oscillation state would be high-amplitude and phase-leading. This
generic non-linear phenomenon, which we refer to as
\textquotedblleft dynamical switching\textquotedblright, is
reminiscent of the usual \textquotedblleft static
switching\textquotedblright\,  of the junction from the
zero-voltage state to the voltage state when the DC current bias exceeds the critical current $I_{0}%
$ \cite{Fulton}. However, an important distinction between
dynamical and static switching is that in dynamical switching, the
phase particle remains confined to only one well of the junction
cosine potential $U(\delta)= -\varphi_{0}I_{0}\cos(\delta)$, and
the time-average value of $\delta$ is always zero. The junction
never switches to the \textquotedblleft normal\textquotedblright\,
state, and thus no DC voltage is generated. Also, for dynamical
switching, the current $i_{c1}$ depends both on $Q$ and on the
detuning $\alpha$. In this Letter we report the first direct
observation of this dynamical switching effect in a Josephson
junction. The parallel with static switching suggests that
dynamical switching can be used for amplification with the added
advantage that no energy dissipation occurs in the junction chip,
a desirable feature for the readout of superconducting qubits
\cite{Vion}.

Typical junction fabrication parameters limit the plasma frequency
to the 20 - 100 GHz range where techniques for addressing junction
dynamics are inconvenient. We have chosen to shunt the junction by
a capacitive admittance to lower the plasma frequency by more than
an order of magnitude and attain a frequency in 1-2 GHz range
(microwave L-band). \ In this frequency range, a simple on-chip
electrodynamic environment with minimum parasitic elements can be
implemented, and the hardware for precise signal generation and
processing is readily available. In our experiment, we directly
measure the plasma resonance in a coherent microwave reflection
measurement. Unlike previous experiments which measured only the
microwave power absorption at the plasma resonance \cite{Dahm,
Pedersen, Holst}, we also measure the phase $\phi$ of the
reflected microwave signal. Thus, we can detect the characteristic
signature of the transition between different oscillating states
of the junction -- a change of oscillation phase relative to the
drive. Note that the phase $\phi$ of the reflected signal which
probes the phase of the oscillation state should not be confused
with the junction gauge-invariant phase difference $\delta$.

\begin{figure}[tbph]
\includegraphics[width=3.1in]{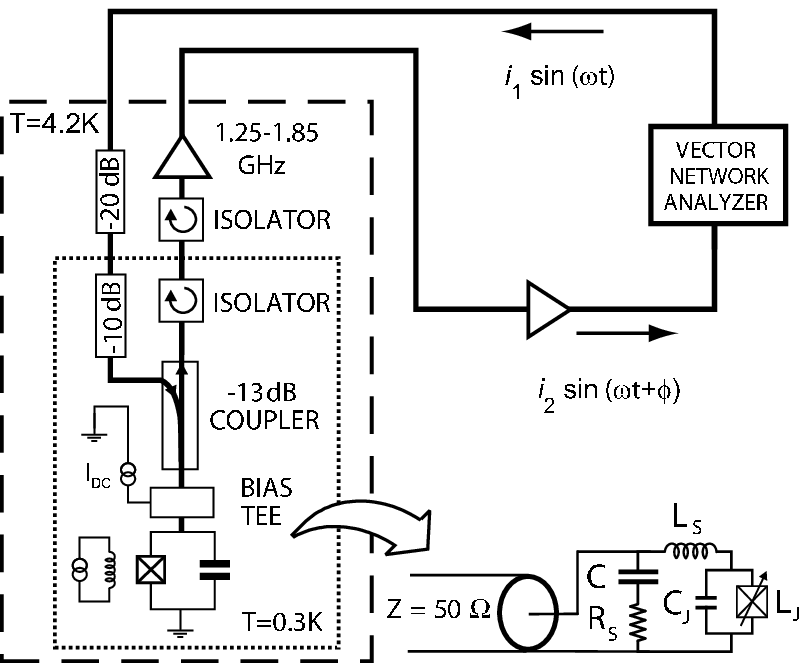} \caption{Schematic of the measurement setup. Thick lines correspond
to $50\,\Omega$ coaxial transmission lines. A lumped element model
for the junction chip and measurement line is shown. The two ideal
current sources actually represent external sources.}
\label{FigSampleWL}
\end{figure}

In the first step of sample fabrication, a metallic underlayer --
either a normal metal (Au, Cu) or a superconductor (Nb) -- was
deposited on a silicon substrate to form one plate of the shunting
capacitor, followed by the deposition of an insulating $\mathrm{Si}%
_{3}\mathrm{N}_{4}$ layer. Using e-beam lithography and
double-angle shadow mask evaporation, we subsequently fabricated
the top capacitor plates along with a micron sized
$\mathrm{Al}/\mathrm{Al}_{2}\mathrm{O}_{3}/\mathrm{Al}$ tunnel
junction. The critical current of the junction\ was in the range
$I_{0\text{ }}=$ $1-2$ $\mathrm{\mu A}$. By varying both the
dielectric layer thickness and the pad area, the capacitance  $C$
was varied between $16$ and $40\,\mathrm{pF}$. Sample parameters
are listed in Table I.

The junction + capacitor chip is placed on a microwave
circuit-board and is wire-bonded to end of a coplanar stripline
which is soldered to a coaxial launcher affixed to the side wall
of the copper sample box. We anchor the RF leak-tight sample box
to the cold stage of a $^{3}\mathrm{He}$ refrigerator with base
temperature $T=280\,\mathrm{mK}$. The measurement setup is
schematically shown in Figure 1. Microwave excitation signals are
generated by a HP 8722D\ vector network analyzer and coupled to
the sample via the -13 dB side port of a directional coupler. The
reflected microwave signal passes through the direct port of the
coupler, and is amplified first using a cryogenic
$1.20-1.85\,\mathrm{GHz}$ HEMT amplifier with noise temperature
$T_{N}=\mathrm{4\,K}$ before returning to the network analyzer. A
DC bias current can be applied to the junction by way of a bias
tee. We use cryogenic attenuators and isolators on the microwave
lines in addition to copper-powder and other passive filters
\cite{Martinis} on the DC lines to shield the junction from
spurious electromagnetic noise.

We locate the linear plasma resonance by sweeping the excitation
frequency from 1 to 2 GHz and measuring the reflection coefficient
$\Gamma(\omega)=i_{2}/i_{1}\,e^{j\phi}=(Z(\omega)-Z_{0})/(Z(\omega)+Z_{0})$,
where $Z_{0}=50\,\Omega$ is the characteristic impedance of our
transmission lines and $Z(\omega)$ is the impedance presented to
the analyzer by the chip and the measurement lines. For an ideal
$LC$ resonator without intrinsic dissipation, we expect a phase
shift $\Delta\phi=\phi_{\omega\gg\omega_{p}}-\phi_{\omega
\ll\omega_{p}}=2\pi$, which we verified by placing a chip
capacitor and an inductive wire bond in place of the junction
chip. An important aspect of our experiment is that $Q$ is now
determined by the ratio $Z_{0}/Z_{J}\sim10$, where
$Z_{J}=\sqrt{L_{J}/(C_{J}+C)}$ and not by the intrinsic junction
losses which are negligible. An excitation power $P=i^{2}Z_{0}/4$
$\thickapprox$ $-120\,\mathrm{dBm}$ $(1\,\mathrm{fW})$
corresponding to a current $i=9\,\mathrm{nA}\ll I_{0}$ keeps the
junction in the linear regime.

\begin{figure}[b]
\includegraphics[width=3.1in]{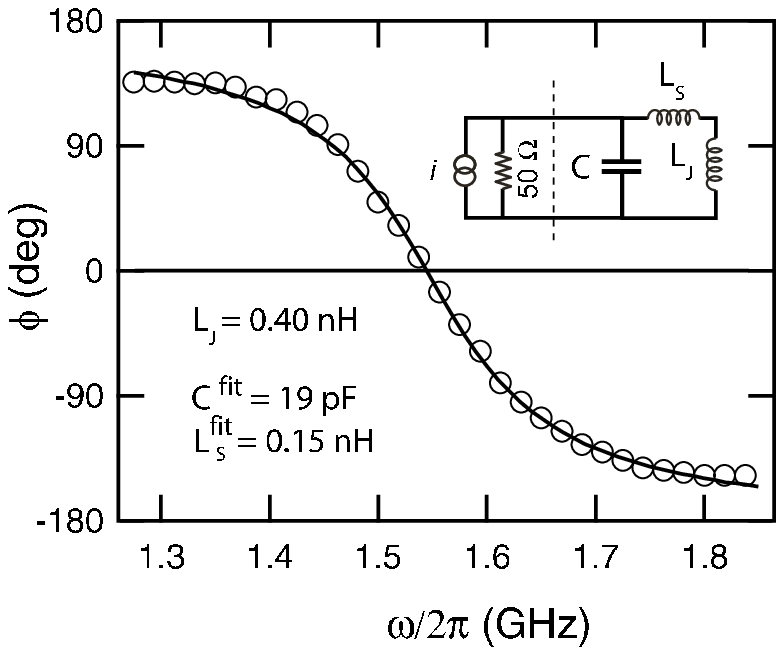} \caption{Normalized reflected signal phase $\phi$ as a
function of excitation frequency for sample 5. The open circles
are measured data for $L_{J}=0.40\,\mathrm{nH}$. The solid line is
calculated from the equivalent circuit model shown in the inset.
The magnitude of the reflected signal is unity.}
\label{FigSampleWL}
\end{figure}

\begin{table}[t]
\begin{tabular}
[c]{|c|c|c||c|c|}\hline
{\small Sample} & $L_{\mathrm{J}}(\mathrm{nH})$ & $\omega_{\mathrm{p}}%
/2\pi(\mathrm{GHz})$ & $C(\mathrm{pF})$ & $L_{\mathrm{S}}(\mathrm{nH}%
)$\\\hline {\small 1} & {\small 0.28} & {\small 1.18} & {\small
39}$\pm1$ & {\small 0.20}$\pm.02$\\\hline {\small 2} & {\small
0.18} & {\small 1.25} & {\small 30}$\pm4$ & {\small
0.34}$\pm.04$\\\hline {\small 2a} & {\small 0.17} & {\small 1.66}
& {\small 18}$\pm1$ & {\small 0.32}$\pm.02$\\\hline {\small 3} &
{\small 0.32} & {\small 1.64} & {\small 16}$\pm1$ & {\small
0.27}$\pm.02$\\\hline {\small 4} & {\small 0.38} & {\small 1.81} &
{\small 19}$\pm1$ & {\small 0.026}$\pm.02$\\\hline {\small 5} &
{\small 0.40} & {\small 1.54} & {\small 19}$\pm1$ & {\small
0.15}$\pm.02$\\\hline
\end{tabular}

\caption{Sample parameters. $L_{J}=\varphi_{0}/I_{0}$ and
$\omega_{\mathrm{p}}$ are measured values. $C$ and
$L_{\mathrm{S}}$ are fit values to the data. Samples 1,2 and 2a
have a 100 nm thick Au underlayer, sample 3 has a 50 nm thick Nb
underlayer, sample 4 has a 1 $\mu\mathrm{m}$ thick Cu underlayer,
and sample 5 has a 200 nm thick Nb underlayer.} \label{tableParam}
\end{table}

In Figure 2, we present the reflected signal phase $\phi$ as a
function of excitation frequency for sample 5. In order to remove
the linear phase evolution associated with the finite length of
the measurement lines, we have subtracted from our measurement in
the superconducting state the reflection coefficient measured with
the junction in the normal state. The point where $\phi=0$ is the
linear-regime plasma frequency. For sample 5,
$\omega_{\mathrm{p}}/\,2\pi=1.54\,\mathrm{GHz}$.

The precise frequency and critical current dependence of the
reflected signal phase of our samples can be accounted for by a
3-element model for the electrodynamic environment seen by the
junction. This lumped element model is shown in the lower right
corner of Figure 1. The parasitic inductance $L_{S}$ and
resistance $R_{S}$ model the non-ideality of the shunting
capacitor $C$. They arise from the imperfect screening of currents
flowing in the capacitor plates and the finite conductivity of
these plates. The plasma frequency in the linear regime is
determined by the total inductance $L_{J}+L_{S}$ and capacitance
$C_{eff}=C_{J}+C\simeq C,$ and is given by the following relation:
\[
\left(  \frac{1}{\omega_{p}}\right)  ^{2}=C(L_{J}+L_{S})=\frac{\varphi_{0}%
C}{I_{0}}+CL_{S}.
\]
We thus plot $1/\omega_{p}^{2}$ versus $1/I_{0}=L_{J}/\varphi_{0}$
in Figure 3 for samples 1, 2, 4 and 5. As the critical current is
decreased by applying a magnetic field, the junction inductance
increases, and the plasma frequency is reduced. For each sample, a
linear fit to the data yields the values of $C$ and $L_{S}$. The
fit values for $C$ agree well with simple estimates made from the
sample geometry. Samples 1 and 2 have nominally the same
capacitance but different critical current, and hence lie
approximately on the same line in Figure 3. A total of four
capacitive pads were used to make the shunting capacitor in
samples 1 and 2, and after initial measurements, we scratched off
two of the pads from sample 2 to obtain sample 2a. The fit
parameters for sample 2a indicate that the capacitance is indeed
halved. For samples with a
thin underlayer (1,2 and 3), a stray inductance in the range $L_{S}%
=0.20-0.34\,\mathrm{nH}$ is observed. For samples 4 and 5 with a
significantly thicker underlayer, $L_{S}$ was reduced to
$0.026\,\mathrm{nH}$ and $0.15\,\mathrm{nH}$ respectively. This
behavior is consistent with the calculated screening properties of
our thin films. To verify that the values of $C$ and $L_{S}$ were
not affected by the magnetic field used to vary $I_{0}$, we varied
$L_{J}$ by applying a DC bias current \cite{Holst} at zero
magnetic field. The resonance data obtained by this method agrees
with the magnetic field data. With a single set of parameters, we
can accurately fit both the position of the resonance and its
lineshape. Additionally, we find $R_{S}=0.8\,\Omega$ for samples 1
and 2, $R_{S}=0.02\,\Omega$ for sample 4, and $R_{S}=0$ for the
superconducting samples 3 and 5. The fit for sample 5 is shown as
a solid line in Figure 2. Finally, we have independently verified
the effect of the shunting capacitor on the plasma resonance by
performing resonant activation experiments \cite{Devoret}.

\begin{figure}[tbpw]
\includegraphics[width=3.1in]{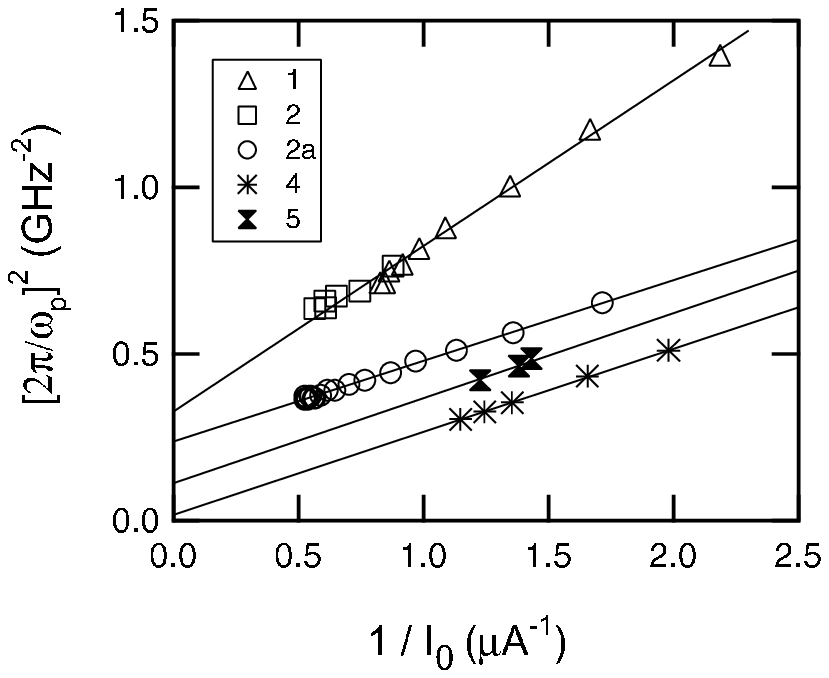} \caption{Inverse square of the plasma frequency $1/\omega_{p}^{2}$
as a function of the inverse critical current
$1/I_{\mathrm{0}}=L_{J}/\varphi_{0}$ for samples 1,2,4 and 5.
Solid lines are linear fits to the data corresponding to the model
of Fig. 1, with a single best fit line drawn for samples 1 and 2
which nominally differ only in $I_{\mathrm{0}}$.}
\label{FigSampleWL}
\end{figure}

In order to study the non-linear regime of the plasma resonance,
we measured the reflection coefficient as a function of frequency
for increasing power. We present the data for sample 5 as a two
dimensional color plot in the lower panel of Figure 4 in which
each row is a single frequency sweep, similar to Figure 2. For
small excitation power, we recover the linear plasma resonance at
$1.54\,\mathrm{GHz}$, shown as a yellow line corresponding to
$\phi=0$. As the power is increased above $-115\,\mathrm{dBm}$,
the plasma frequency decreases, as is expected for a junction
driven with large amplitude \cite{Devoret}. The boundary between
the leading-phase region (green) and the lagging-phase region
(red) therefore curves for high powers. This curvature has an
interesting consequence: When we increase the drive power at a
constant frequency slightly below the plasma frequency, the phase
as a function of power undergoes an abrupt step, as predicted
\cite{Dykman}. For yet greater powers $(>-90\,\mathrm{dBm})$, we
encounter a new dynamical regime (black region in Figure 4) where
$\delta$ appears to diffuse between the wells of the cosine
potential. When measuring the AC junction resistance at
$f=221\,\mathrm{Hz}$ in presence of the microwave drive, a finite
resistance was only observed in the black region. To verify that
this phenomenon was not due to junction heating, we have varied
the ramp time of the power and no change in the transition power
was observed. In the lower panel of Figure 4, we illustrate the
sequence of dynamical transitions by plotting $\phi$ as a function
of incident power at $\omega/2\pi =1.375\,\mathrm{GHz}$. For $P
<-102\,\mathrm{dBm}$, the phase is independent of power ($\delta$
oscillates in a single well in the harmonic-like, phase-leading
state, letter A). For
$-102\,\mathrm{dBm\,}<\,P<-90\,\mathrm{dBm}$, the phase evolves
with power and $\delta$ still remains within the same well, but
oscillates in the anharmonic phase-lagging state (letter B).
Finally, for $P>-90\,\mathrm{dBm}$, the average phase of the
reflected signal saturates to -180 degrees, corresponding to a
capacitive short circuit (letter C). This last value is expected
if $\delta$ hops randomly between wells, the effect of which is to
neutralize the Josephson inductive admittance.

\begin{figure}[tbpw]
\includegraphics[width=3.1in]{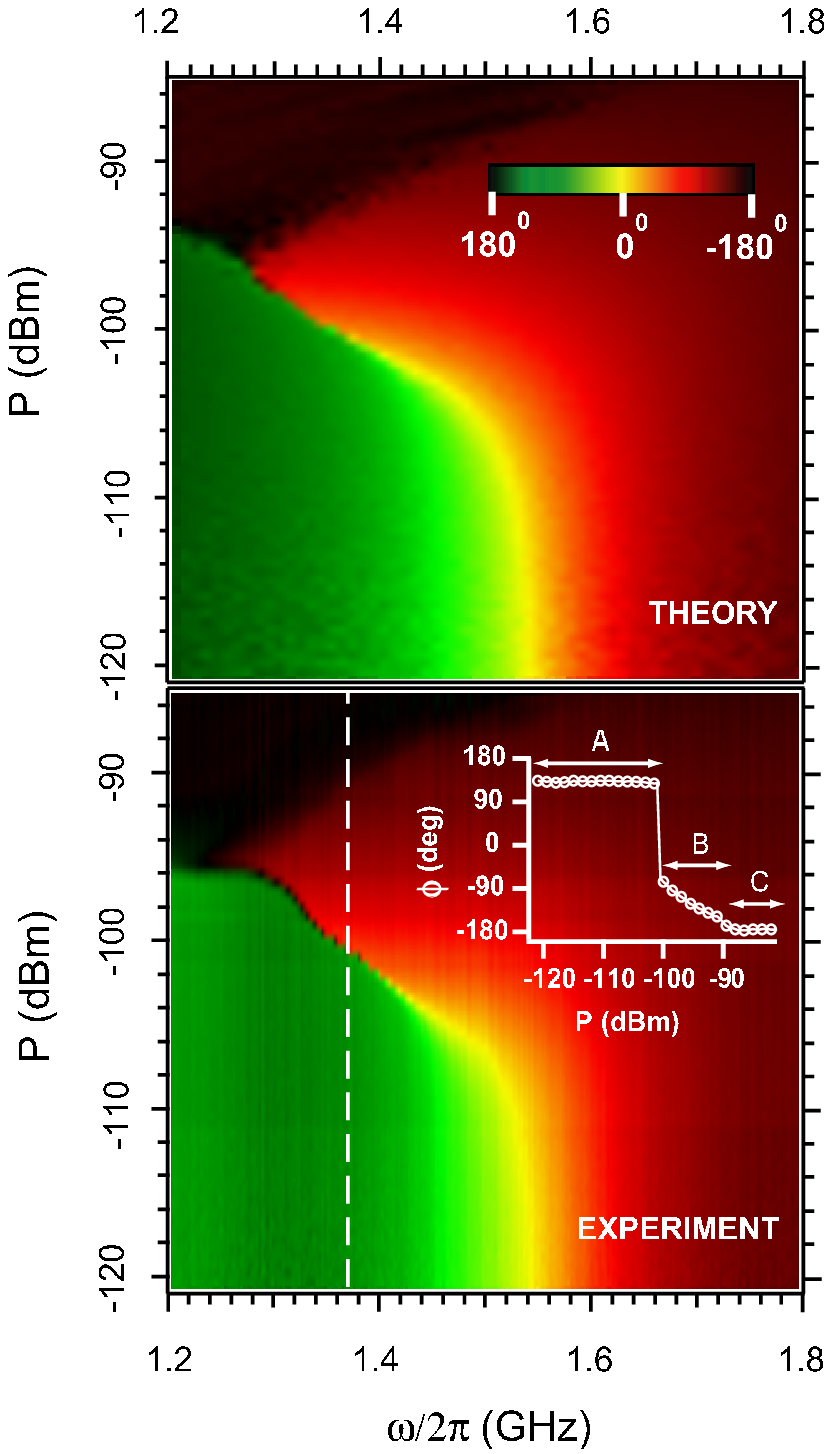} \caption{Normalized reflected signal phase $\phi$
(wrap-around color scheme) as a function of excitation frequency
$\omega/2\pi$ and excitation power $P$. In the lower panel, a
vertical slice taken at $\omega/2\pi=1.375\,\mathrm{GHz}$ (dashed
line) shows the abrupt transition between two oscillation states
of the system. The upper panel is the result of numerical
simulations.} \label{FigSampleWL}
\end{figure}

The value of the switching current $i_{c1}$ for the A-B
transition, which is a function of both the detuning $\alpha$ and
power $P$, is in good qualitative agreement with the analytical
theory which retains only the first anharmonic term in the cosine
potential \cite{Dykman}. For instance, the slope of the A-B
transition line at the linecut in Figure 4,
$dP(\mathrm{dBm})/d\alpha(\%)=0.8$ for the experiment while we
calculate its value to be $0.7$. Furthermore, in measurements in
which the power is ramped in less than $100\,\mathrm{ns}$, we
verified that the transition between dynamical states is
hysteretic, another prediction of the theory. These results will
be presented in a later publication. To explain the complete
frequency and power dependence of the transitions shown in the
lower panel of Figure 4, we have performed numerical simulations
by solving the full circuit model of the lower corner of Figure 1,
including the exact junction non-linear constitutive relation. The
result of this calculation is shown in the upper panel of Figure
4. It correctly predicts the variation of the plasma frequency
with excitation power, and the boundaries of the phase diffusion
region. The agreement between theory and experiment is remarkable
in view of the simplicity of the model used with no adjustable
parameters, and only small differences in exact shape of region
boundaries are observed. It is important to mention that the
overall topology of Figure 4 is unaffected by changes in the
parameter values within the bounds of Table I.

In conclusion, we have performed a novel, phase-sensitive,
microwave experiment demonstrating that the Josephson plasma
oscillation can switch between the two dynamical states predicted
for a driven non-linear system. At the critical excitation power
we observe an abrupt change in the reflected signal phase. In
accordance with the analytical theory and numerical simulations,
we have observed that this critical power is a strong function of
the junction critical current. This phenomenon can therefore be
applied to the detection of small relative variations in critical
current in the same manner that the usual DC SQUID, biased in the
vicinity of the critical current, transforms a flux-induced
variation in critical current into a DC voltage \cite{Clarke}.
Furthermore, following the methodology invented for experiments
with trapped electron systems \cite{Gabrielse}, we can use this
dynamical switching as the basis for a single-shot and latching
qubit readout. Since the measurement of phase is purely
dispersive, this new readout would have advantage the of
eliminating dissipation at the chip level, a limiting factor for
superconducting qubits.

We would like to thank D. Prober for use of his laboratory
equipment and useful discussions, Abdel Aassime for help in the
initial sample fabrication, and L. Grober for assistance with the
electron microscope. This work was supported by the ARDA (ARO
Grant DAAD19-02-1-0044) and the NSF (Grant DMR-0072022).

\end{document}